\documentstyle[prl,aps,epsfig]{revtex}
\twocolumn
\begin{document}
\draft
\twocolumn[\hsize\textwidth\columnwidth\hsize\csname@twocolumnfalse\endcsname
\title{Highly localized clustering states in a granular gas driven 
by a vibrating wall}
\author{Eli Livne $^{1}$, Baruch Meerson $^{1}$
and Pavel V. Sasorov $^{2}$}
\address{$^{1}$ Racah Institute of Physics, Hebrew
University   of  Jerusalem, Jerusalem 91904, Israel}
\address{$^{2}$ Institute of Theoretical and Experimental Physics, Moscow
117259, Russia}
\maketitle
\begin{abstract}
An ensemble of inelastically colliding grains driven by a
vibrating wall in 2D exhibits density clustering.
Working in the limit of  nearly elastic collisions and employing
granular hydrodynamics, we predict, by a marginal stability
analysis, a spontaneous symmetry breaking of the extended
clustering state (ECS). 2D steady-state solutions
found numerically describe localized clustering states (LCSs). 
A time-dependent granular hydrodynamic simulation
shows that LCSs can develop from natural initial
conditions. The predicted instability should be observable
in experiment.
\end{abstract}
 \vskip1pc]
\narrowtext

Granular flows exhibit fascinating non-equilibrium phenomena and
continue to attract much interest \cite{Jaeger,Kadanoff}. We will
concentrate here on the striking tendency of granular ``gases"
(rapid granular flows) to form dense clusters \cite{Hopkins}.
Clustering results from energy losses by inelastic collisions, and
it is a manifestation of thermal condensation instability
\cite{plasma}. Since the discovery of the clustering instability,
the validity of granular hydrodynamics \cite{Campbell} has been under
scrutiny. In a freely cooling granular gas,  all grains eventually
come to rest, making the hydrodynamic  (and even kinetic)
description problematic. In a {\it driven} granular gas 
hydrodynamics can be conveniently tested on its steady
states. The simplest system of this kind is a 
submonolayer of
grains in 2D, driven by a vibrating side wall at
zero gravity. This and related ``test bed'' systems were
investigated by molecular dynamic (MD) simulations
\cite{Grossman,Brey,Esipov} and in experiment \cite{Kudrolli}.
For sufficiently high average densities, 
ECS, was observed in these works apart from the driving wall.
The basic physics of the ECS is simple. Because
of
the inelastic collisions the
granular temperature decreases with the distance from the driving
wall. To
maintain the momentum balance, the granular density 
increases with this distance, reaching the maximum
at the opposite (``elastic'') wall. When the density contrast is 
large enough, the enhanced density 
region is observed as the 
ECS.

Comparisons of the steady-state density profiles obtained in MD
simulations of this problem
with those
predicted by granular hydrodynamics showed that hydrodynamics is
valid only in the limit of nearly elastic collisions
\cite{Grossman,Brey,Esipov,hydro}. This limit  
has not
been fully explored, and it is non-trivial. 
To show it, we perform a
stability
analysis of this simple
granular system.
This analysis reveals a symmetry-breaking instability of the ECS,
and formation of LCSs. This result puts this system in the
list of pattern-forming systems out of equilibrium \cite{Cross}.

Consider a big ensemble of identical spherical grains of
diameter $d$ and mass $m_g=1$ rolling on a smooth horizontal surface
of a rectangular box with dimensions $L_x \times L_y$. The number
density of grains is $n (x,y)$. For a submonolayer coverage the
maximum value of $n$ corresponds to the (hexagonal) close-packing
value $n_c = 2/(\sqrt{3} d^2)$. Three of the walls are immobile, and
grain collisions with
them are assumed
elastic. The fourth wall
(located at $x=L_x$) supplies energy to the granulate. We will consider
two different models of energy supply, see below.
The energy
is being lost through inelastic hard-core grain collisions. We
neglect the grain rotation and parameterize the inelasticity of
grain collisions by a constant normal restitution coefficient $r$.

We assume a {\it strong} inequality $1-r^2 \ll 1$, which 
makes a 
hydrodynamic description 
valid \cite{Grossman,Brey,Esipov}. Therefore, steady 
states of the system can be described
by the equations of
momentum and energy balance:
\begin{equation}
p=const\,,\,\,\, \nabla \cdot (\kappa \, \nabla T) = I \,,
\label{energy1}
\end{equation}
where $p$ is the granular pressure, $\kappa$ is the thermal
conductivity, $I$ is the rate of energy losses by collisions and
$T$ is the granular temperature. To proceed, one needs an equation of state
$p=p\,(n,T)$ and relations for $\kappa$ and $I$ in terms of $n$
and $T$. In the low-density limit, $n \ll n_c$, these relations
can be derived from the Boltzmann equation \cite{Campbell}. 
The high-density limit, $n_c-n \ll n_c$, was
considered by Grossman {\it et al.} \cite{Grossman}. They also
suggested convenient 
interpolations between the low- and high-density limits, and verified
them by a detailed comparison with MD simulations. We will
adopt this practical approach (see,
however, Ref. \cite{robust}).
In our notation
\begin{equation}
p= nT\, \frac{n_c+n}{n_c-n}\,, \label{state}
\end{equation}
$\kappa=(\mu/l)\, n\,(\alpha l+d)^2\,T^{1/2}$ and
$I=(\mu/\gamma\,l)\,(1-r^2)\,n\,T^{3/2}$. Here $l$ is the
mean free path of the grains,
\begin{equation}
l=\frac{1}{\sqrt{8}nd}\, \frac{n_c-n}{n_c-an}\,,
\label{meanfree}
\end{equation}
$a= 1-(3/8)^{1/2}$, and $\alpha$ and $\gamma$ are numerical
factors of order unity. Grossman {\it et al.} \cite{Grossman}
found that $\alpha \simeq 1.15$ and $\gamma \simeq 2.26$. 
The value of $\mu$, another numerical factor of order unity, is
irrelevant in the steady-state problem.

The boundary conditions 
include the no-flux conditions $\nabla_n T = 0$ at the
``elastic'' walls $x=0,\, y=0$ and $y=L_y$. Previously,
the ``thermal'' wall condition $T=const$ was used at $x=L_x$
\cite{Grossman,Brey,Esipov,twowalls}. We will
use a different boundary condition, to simulate the vibrating
wall more directly. Our main results, however, will be shown to
hold for the
thermal wall as well.

The problem of computing the energy flux $q$ from a vibrating wall
into granulate 
has been addressed in several works
\cite{McNamara2}. Let the wall oscillate 
sinusoidally: $x=L_x + A \cos \omega t$. For small
area fractions
the granulate near this wall is in the dilute limit.
We assume $A \ll l$, so the vibrating wall does not generate
any collective motions 
in the granulate. Grain collisions with the vibrating wall are 
assumed elastic. Also, 
$\omega$ is much larger than the rate of granular collisions
near the vibrating wall,
$T^{1/2}/l$, so there are no
correlations
between two
successive grain collisions with the wall. 
The limit $A \omega \ll T^{1/2}$ was
considered by Kumaran
\cite{Kumaran} for a non-zero gravity. A direct calculation 
analogous to his, but for zero gravity, yields $q =
(2/\pi)^{1/2}\, A^2 \, \omega^2\,n \,T^{1/2}$ \cite{interpolation}. In the
language of hydrodynamics, $q$ is the heat flux at the wall:
\begin{equation}
\kappa \, \partial T/\partial x = q\,\,\, \mbox{at} \,\, x=L_x\,.
\label{wall1}
\end{equation}
Finally, $\int_0^{L_x}\int_0^{L_y}\, n(x,y)\, dx \, dy= \langle n \rangle\,
L_x \,L_y$ is normalization condition, 
where $\langle n \rangle$ is the average grain density.

Using Eq. (\ref{state}), we eliminate $T$ in favor of $n$ and $p$.
In its turn, $p$ can be eliminated by integrating Eq.
(\ref{energy1}) over the whole box and using the Gauss theorem and
Eq. (\ref{wall1}). It is convenient to write the governing
equations in a scaled form. Introduce scaled coordinates:  ${\bf
r} /L_x \to {\bf r}$ so that the box dimensions become $1 \times
\Delta$, where $\Delta = L_y/L_x$ is the aspect ratio of the box.
Introducing the (scaled) inverse granular density
$z\,(x,y)=n_c/n\,(x,y)$, we obtain
\begin{equation}
\nabla \cdot \left( F(z) \nabla z \right) ={\cal L}\, Q(z)\,.
\label{energy2}
\end{equation}
The boundary conditions are
\begin{equation}
\nabla_n z = 0 \,\,\, \mbox{at} \,\,\, x=0\,,\, y=
0\,\,\,\mbox{and}\,\,\, y=\Delta\,,
\label{walls}
\end{equation}
and
\begin{equation}
\left.\left(G(z)\frac{\partial z}{\partial
x}\right)\right|_{x=1}={\cal L}\,
\frac{\int\limits_0^1\int\limits_0^\Delta\, Q\, dx\, dy}
{\int\limits_0^\Delta\, H\left[z\,(1,\,y)\right]\, dy}\,.
\label{wall2}
\end{equation}
The normalization condition becomes
\begin{equation}
\int_0^1\,\int_0^{\Delta} \, z^{-1}\,dx\,dy= f\,\Delta\,,
\label{normalization}
\end{equation}
while functions $F,G,H$ and $Q$ are the following:
\begin{equation}
F(z)=\frac{(z^2+2z-1)[\alpha z(z-1)+\sqrt{32/3}(z-a)]^2}
{(z-a)(z-1)^{1/2}z^{3/2}(z+1)^{5/2}}\,,
\label{F}
\end{equation}
\begin{equation}
G(z)=\frac{(z^2+2z-1) [\alpha
z(z-1)+\sqrt{32/3}(z-a)]^2}{z(z-a)(z-1)(z+1)^2}\,,
\label{G}
\end{equation}

\begin{equation}
H(z) = \frac{F(z)}{G(z)}\,\,\,\,\mbox{and}\,\,\,\,Q(z)=\frac{(z-a)
(z-1)^{1/2}}{(z+1)^{3/2} z^{1/2}}\,. \label{Q}
\end{equation}
Finally,
${\cal L} = (32/3\gamma)\,(L_x/d)^2\,(1-r^2)$. The other
two governing parameters are the grain area fraction $f=\langle n
\rangle /n_c$ and $\Delta$. Notice that the steady-state 
{\it density} distributions are independent
of $A$ and $\omega$. 

Eqs. (\ref{energy2})-(\ref{normalization}) make a closed set.
Their 1D ($y$-independent) solution $z = Z(x)$ is described by
equations
\begin {equation}
\left(F Z^{\prime}\right)^{\prime} = {\cal L} Q\,,\,\,\,\,\,
\left.Z^{\prime}  \right|_{x=0} = 0 \,\,\, \mbox {and}
\int\limits_0^1\, Z^{-1}\, dx = f\,,
\label{1d}
\end{equation}
where primes stand for the $x$-derivatives. Eq.
(\ref{wall2}) is now satisfied automatically. Eqs. (\ref{1d})
coincide with those obtained by Grossman {\it et al.}
\cite{Grossman} for a {\it thermal} wall at $x=1$. Therefore,
the {\it density} profiles of the 1D states coincide for the
different types of driving.
Eqs. (\ref{1d}) can be solved analytically in the high- and
low-density limits \cite{Grossman}. These solutions clearly show
that criterion \cite{hydro} for validity of hydrodynamics is
equivalent to a strong inequality $1-r^2 \ll 1$.

Most interesting among the 1D states 
is the state with a {\it dense} cluster (ECS) located
at the elastic wall $x=0$, and a low-density region elsewhere. 
In this case
Eqs. (\ref{1d}) should be solved numerically. Examples are
presented in Ref. \cite{Grossman}, and a similar clustering state 
(CS) was observed
experimentally \cite{Kudrolli}. The main objective of this Letter is
to show that this state (uniform in the $y$-direction) can give way,
via a spontaneous symmetry breaking, to CSs highly
localized in the $y$-direction. First, a marginal stability (MS)
analysis 
will show, in some region of parameters,
loss of stability of the ECS . Then, solving Eqs.
(\ref{energy2})-(\ref{normalization}) numerically, we will find
LCSs.  Finally,  a
time-dependent granular hydrodynamic simulation will show that
highly localized CSs
can develop from natural initial conditions.

Linearizing Eqs. (\ref{energy2})-(\ref{normalization}) around  the
ECS $z=Z(x)$ and looking for a small correction in the
form of $\psi_m(x) \cos \,(\pi m y/\Delta)$, where $m=1,\,2,
\dots\,$, we obtain
\begin{equation}
F \phi^{\prime\prime}-\left( {\cal L}\, Q_Z \, +\pi^2 m^2
\Delta^{-2}\,F \right) \phi = 0\,,
\label{phi}
\end{equation}
where $\phi = F \psi_m$ and index $Z$ means the $z$-derivative
evaluated at $z=Z (x)$. The boundary conditions are
\begin{equation}
\phi^{\prime} |_{x=0} = 0\,,\,\, \left.\left(F G \phi^{\prime} +
Z^{\prime} (F G_Z - G F_Z) \phi \right) \right|_{x=1} = 0\,.
\label{bound2}
\end{equation}
Functions $F$ and $G$ which enter Eqs. (\ref{phi}) and
(\ref{bound2}) are evaluated at $z=Z(x)$.

For fixed values of ${\cal L}$, $f$ and $m$, Eqs.
(\ref{phi}) and (\ref{bound2}) represent a linear eigenvalue problem for
$\Delta^{-2}$, or simply for the aspect ratio
$\Delta$. We denote these eigenvalues by
$\Delta_m$, $m=1,\,2\, \dots$. Obviously, $\Delta_m ({\cal L},f)=
m \Delta_1 ({\cal L},f)$. 
The eigenvalues $\Delta_m$ (when they exist) 
represent the {\it critical} values of the aspect ratio:
at $\Delta>\Delta_m$ the ECS looses stability.  
For the mode $m=1$ (or $\lambda/2$, one half of the wavelength across 
the system in the $y$-direction) the critical value $\Delta_1$ is the 
lowest. Figure 1 shows, for
different values of ${\cal L}$, the MS curves $\Delta=\Delta_1
(f)$ that we found numerically. Above the MS curves LCSs must develop.
Interestingly, the ECS remains linearly stable {\it for
any} $\Delta$ beyond a finite interval of the area fractions $f_1
({\cal L}) <f< f_2({\cal L})$ such that $f_1 > 0$ and $f_2<1$. As
${\cal L}$ increases, the instability interval $(f_1, f_2)$
shrinks, while the minimum value of $\Delta_1$ decreases:
$\Delta_1^{(min)} \simeq 140\,{\cal L}^{-1/2}$ \cite{delta}. For
sufficiently large ${\cal L}$, $\Delta_1^{(min)}$ becomes less
than 1. See Fig. 2 which shows the instability tongues $m=1$ and
$m=2$ for ${\cal L}=5 \cdot 10^4$. For higher $m$
we obtain modes $3 \lambda/2,\,
2 \lambda, \, 5 \lambda/2, \, \dots$ which fit in boxes with
increasingly
larger aspect
ratios, $\Delta > m \Delta_1$.

\begin{figure}[h]
\centerline{ \epsfig{file=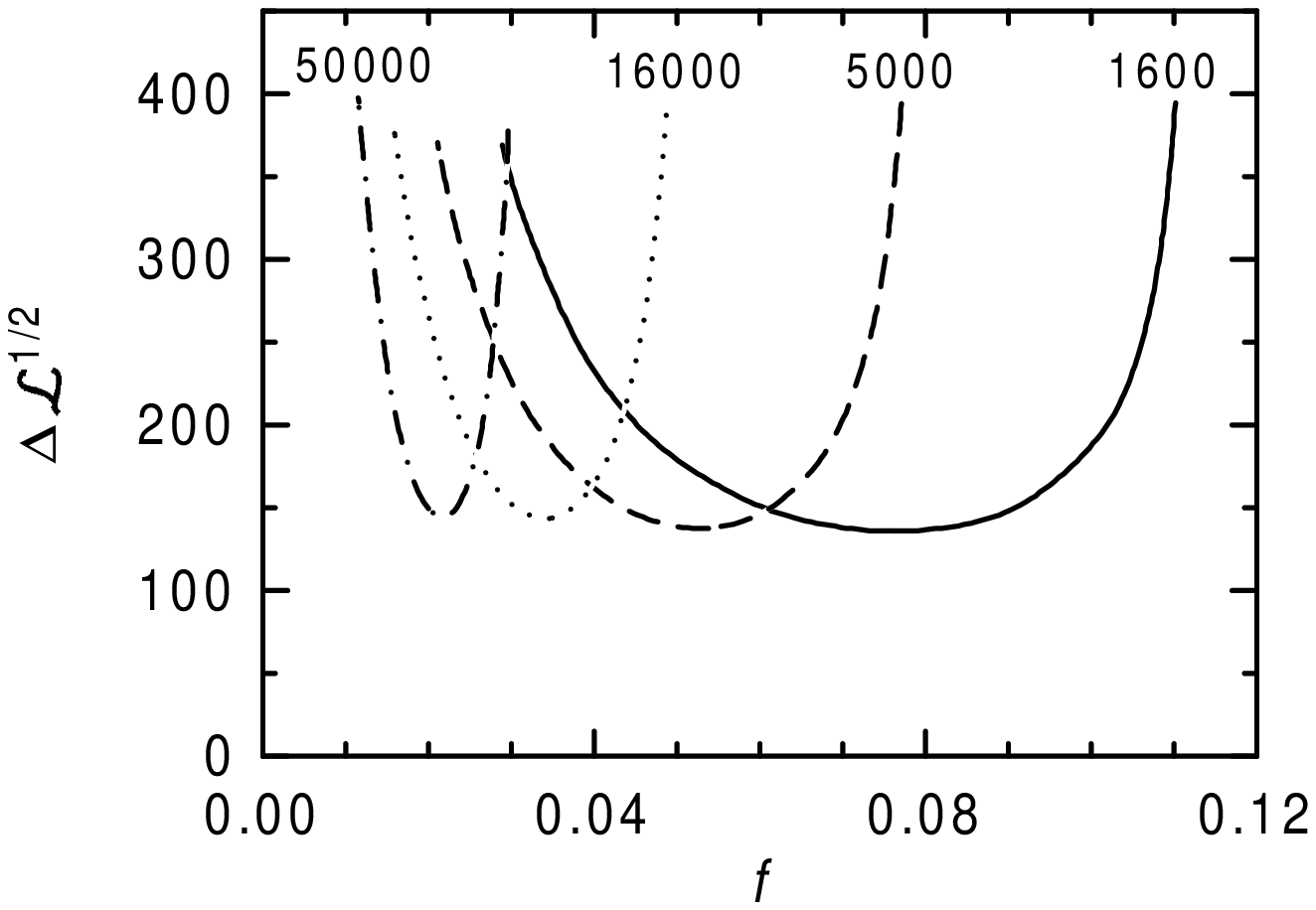, width=2.4in, clip= }}
\caption{Marginal stability curves $\Delta_1 (f)$ for different
values of ${\cal L}$. The values of $\Delta_1$ are multiplied by
${\cal L}^{1/2}$.} \label{Fig. 1}
\end{figure}

\begin{figure}[h]
\centerline{ \epsfig{file=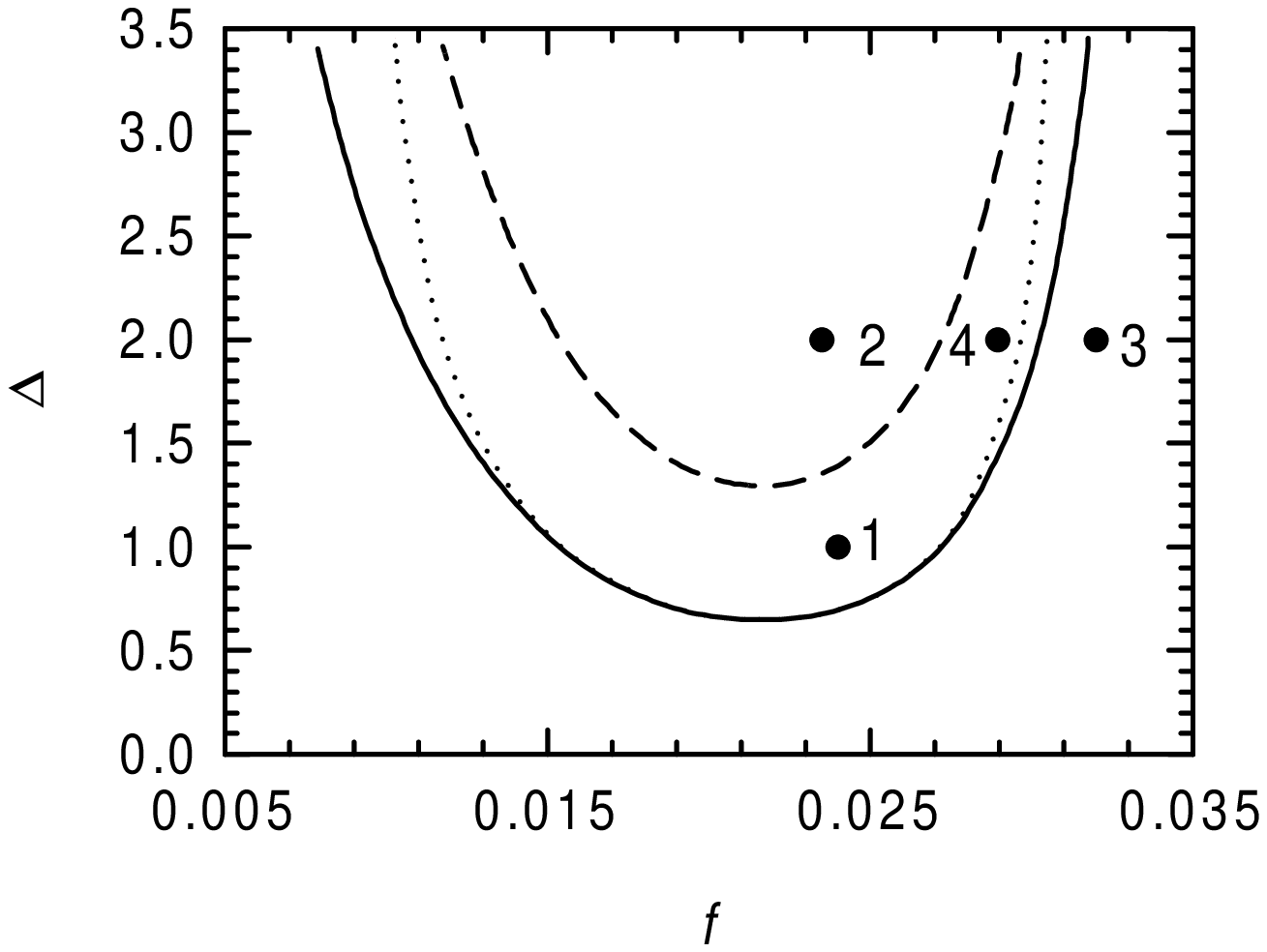, width=2.4in, clip= }}
\vspace{0.1in}\caption{Marginal stability (MS) curves $\Delta_m
(f)$ for $m=1$ (solid line) and $2$ (dashed  line) and ${\cal
L}=5\cdot 10^4$. Density profiles corresponding to points 1,2,3
and 4 are shown in Fig. 3. The dotted line shows the MS curve
$m=1$ for the ``thermal wall".} \label{Fig. 2}
\end{figure}

When $f \ll \min \, (1,\, {\cal L}^{-1/2})\,$, 
$\,\Delta_1(f)$ can be found analytically. In this case the  whole
system is in the dilute limit, $z \gg 1$ (still, it is necessary
to account for the subleading terms). In addition, $Z(1)-Z(0) \ll
Z(0)$ in this case, so Taylor expansion of $Z (x)$ and $\psi (x)$ up
to $x^4$ suffice. After some algebra, Eqs. (\ref{1d}) -
(\ref{bound2}) yield
\begin{equation}
\Delta_1 = \pi \left( \frac{{\cal L}^2 f^4}{3 \alpha^4} -
\frac{(1+a)\, {\cal L} \, f^3}{\alpha^2} \right)^{-1/2}\,.
\label{asymp}
\end{equation}
It follows from Eq. (\ref{asymp}) that $f_1 ({\cal L}) = 3\,
\alpha^2 (1+a) \,{\cal L}^{-1}$. 

The same instability appears when the wall $x=1$ (in scaled units)
is ``thermal''. Solving the corresponding eigenvalue problem
[where the second boundary condition in Eq. (\ref{bound2}) is
replaced by $\phi(x=1) = 0$], we obtained instability tongues
similar to those for the vibrating wall, but more narrow. Figure 2
shows the instability tongue $m=1$ for ${\cal L}= 5 \cdot 10^4$.
Noticeable is the coincidence of the $m=1$ curves at intermediate
$f$ for the two types of driving. It results from a strong
localization of the eigenfunction $\phi(x)$ near the elastic
wall $x=0$ at
large ${\cal L}$ and intermediate $f$. The exact form of the 
boundary condition at
$x=1$ is not important in this regime. Finally, for the thermal
wall the LCS is stable for any $\Delta$ if $f \ll \min \,
(1,\, {\cal L}^{-1/2})$, in contrast to the vibrating wall.

In the rest of the paper we will deal with the vibrating wall.
Within the instability tongues of Fig. 1 the MS analysis is
invalid. Besides, it can miss a subcritical bifurcation outside of
the instability tongues. Therefore, we solved the 2D
steady-state equations (\ref{energy2})-(\ref{normalization})
numerically (a nonlinear Poisson solver, Newton's iterations),
exploring some parts of the parameter plane $(f,\,\Delta)$ of Fig.
2. Figure 3 shows the density profiles of 4 typical steady states
with $m=1$ and $2$.
Highly localized nonlinear $\lambda/2$- and $\lambda$-states are
evident in Fig. 3a and b. The maximum/minimum density ratio
along the elastic wall $x=0$ reaches about $21$ in these examples.

\begin{figure}[h]
\centerline{ \epsfig{file=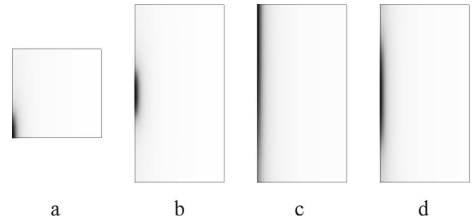, width=2.4in, clip= }}
\vspace{0.1in} \caption{Steady-state density profiles (gray scale,
separate for each picture) corresponding to points 1 (a), 2 (b), 3
(c) and 4 (d) of Fig. 2. The maximum (minimum) density  at the
wall $x=0$ is 0.76 (0.036) (a and b), 0.48 (0.21) (c) and 0.54
(0.10) (d). The gas density at the vibrating wall $x=1$ is close
to $4 \cdot 10^{-3}$ for all profiles.} \label{Fig. 3}
\end{figure}

In general, we found that when crossing the MS curve
$\Delta=\Delta_1 (f)$ from the left (along the line $\Delta=2$),
or from below, one goes continuously from an ECS to a
``weakly 2D" $\lambda/2$-state. This implies a
supercritical bifurcation. However, when moving from the right to
the left along the line $\Delta=2$, nonlinear $\lambda/2$- and
$\lambda$-states appear {\it inside} the linear stability regions
of the ECS and of the mode $m=1$, respectively, and {\it
coexist} with the ECS (with the mode $m=1$, respectively).
These findings give evidence for bistability and subcritical
bifurcations. Examples of subcritical $\lambda/2$- and
$\lambda$-states are shown in Fig. 3c and d. Super- and subcritical 
LCSs were also observed for $\Delta=3$. 

A mirror 
reflection of  Fig. 3a with respect to $y=0$ makes 
$\Delta=2$ and
produces a 
$\lambda$-state similar to Fig. 2b. Furthermore, extending Fig. 2b
periodically in the $y$-direction, we obtain a periodic 
chain of LCSs which fit in boxes with increasingly larger
aspect ratios. When the aspect ratio goes to infinity,
the periodic cluster chain becomes infinite. 
Cluster chains with {\it different} 
periods can fit in boxes with large enough aspect ratios,
therefore, an
interesting
selection problem appears, like in other 
pattern-forming systems
\cite{Cross}. 

We performed a series
of {\it
time-dependent} hydrodynamic simulations (with $\Delta=1,2$ and $3$)
which showed that the LCSs are 
dynamically stable and develop from natural
initial conditions. We will briefly report
here a single simulation with $\Delta=2$. The full hydrodynamic equations 
were solved with the
same constitutive relations and boundary conditions as in the steady
state analysis.  Instead of the shear viscosity in the momentum
equation we
accounted for a small rolling friction force $- n {\bf v}/{\tau}$. An
extended version of the compressible hydro 
code VULCAN \cite{vulcan} was used. 

The initial scaled density in this example was
$n(x,y,t=0) = f + 0.1\,f\,\cos (\pi y)$ (independent of
$x$). Figure 4 shows the density evolution. A cluster
develops near the elastic wall $x=0$. With 
time it becomes
localized in the $y$-direction and approaches the
steady-state profile shown in Fig. 3b. 
\begin{figure}[h]
\centerline{ \epsfig{file=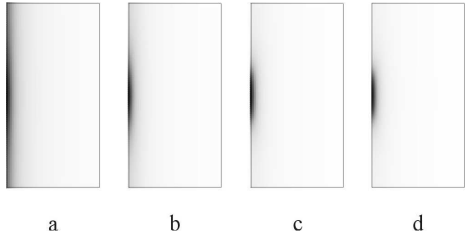, width=2.4in, clip= }}
\vspace{0.1in} \caption{Density evolution for ${\cal L} = 5\cdot
10^4,\, \Delta =2$ and $f=0.0235$. Shown are the density profiles
(gray scale, separate for each picture) at scaled times 100 (a),
500 (b), 1,000 (c) and 1,290(d). The maximum (minimum) density at
the wall $x=0$ is 0.25 (0.14) (a), 0.46 (0.072) (b), 0.66 (0.040)
(c) and 0.74 (0.036) (d). The gas density at the vibrating wall
$x=1$ is close to $4 \cdot 10^{-3}$ for all profiles.} \label{Fig.
4}
\end{figure}
In summary, we predict a spontaneous transition from an extended
to highly localized clustering states 
in a driven submonolayer granular system. The 
transition should occur
when the aspect ratio is large enough (see Fig. 1). It
is insensitive to the vibration frequency and amplitude, and only
weakly depends on the type of the driving wall. To observe
this transition in experiment, one should
minimize the role of the
rolling/sliding friction \cite{Kudrolli}, unaccounted for 
in our model.
The frictional energy losses are proportional to $T$, while
the collisional energy losses
are proportional to $T^{3/2}$. Therefore, one should
work with higher granular temperatures (that is, larger $A \omega$).

When  
$1-r^2$ is {\it not} small,
the normal stress difference, non-Gaussianity in the velocity
distribution
and possible
lack of scale separation become important. The role of these
effects in the symmetry-breaking instability 
should be the subject of further studies.

Finally,
the aspect ratios used in Refs. \cite{Grossman,Brey,Kudrolli} were
always lower than the threshold values for the instability 
$\Delta_1^{(min)}$. As the result, the instability
was suppressed by granular heat conduction in 
the $y$ direction.
 
We acknowledge useful discussions with E. Ben-Naim, 
J.
Fineberg and J.P. Gollub. This research was supported by the Israel Science
Foundation founded by the Israel Academy of Sciences and
Humanities, and by the Russian Foundation for Basic Research
(grant No. 99-01-00123).


\begin{references}
\bibitem{hydro} Hydrodynamics is expected to be valid when
the mean free path of the grains is much smaller than any length
scale (and the mean collision time is much smaller than any time
scale) described hydrodynamically.
\bibitem{robust} Notice that
the
symmetry-breaking instability predicted in this work 
does not require very special  
constitutive relations (CRs). It
appears already in the dilute limit described 
by the ``standard'' CRs \cite{Campbell}.
\bibitem{interpolation}  In the opposite limit $T^{1/2} \ll A \, \omega$ 
we found $q = 2 \pi A \omega n T$. An interpolation formula
$q= 2\pi A \omega \,n T\,(1+2^{1/2} \pi^{3/2} \epsilon)^{-1}$
should work well for any $\epsilon= T^{1/2}/(A \omega)$.
\end{references}
\end{document}